\documentclass{aa}

\usepackage{graphicx}
\usepackage{txfonts}
\usepackage{natbib}
\usepackage[colorlinks=true,citecolor=blue]{hyperref}
\usepackage{xspace}
\usepackage{array}
\usepackage[labelformat=simple]{caption,subcaption}
\usepackage{gensymb}

\bibpunct{(}{)}{;}{a}{}{,}

\newcommand{\Teff}{\ensuremath{T_\mathrm{eff}}\xspace}
\newcommand{\logg}{\ensuremath{\log g}\xspace}

\usepackage{color}

\definecolor{teal}{rgb}{0.0, 0.5, 0.5}

\begin{document}

\title{Direct imaging and spectroscopy of exoplanets with the ELT/HARMONI high-contrast module}

\author{
    M.~Houllé\inst{\ref{lam}, \ref{eso}}
    \and
    A.~Vigan\inst{\ref{lam}}
    \and
    A.~Carlotti\inst{\ref{ipag}}
    \and
    É.~Choquet\inst{\ref{lam}}
    \and
    F.~Cantalloube\inst{\ref{mpia}}
    \and
    M.~W.~Phillips\inst{\ref{exeter}}
    \and
    J.-F.~Sauvage\inst{\ref{lam},\ref{onera}}
    \and
    N.~Schwartz\inst{\ref{ukatc}}
    \and
    G.~P.~P.~L.~Otten\inst{\ref{lam}}
    \and
    I.~Baraffe\inst{\ref{exeter},\ref{enslyon}}
    \and
    A.~Emsenhuber\inst{\ref{bern},\ref{tucson}}
    \and
    C.~Mordasini\inst{\ref{bern}}
}

\institute{
    Aix Marseille Univ, CNRS, CNES, LAM, Marseille, France\\
    \email{mathis.houlle@lam.fr}\label{lam}
    \and
    European Southern Observatory, Alonso de Córdova 3107, Santiago, Chile \label{eso}
    \and
    Univ. Grenoble Alpes, CNRS, IPAG, 38000 Grenoble, France \label{ipag}
    \and
    Max Planck Institute for Astronomy, Königstuhl 17, 69117, Heidelberg, Germany \label{mpia}
    \and
    School of Physics and Astronomy, University of Exeter, Exeter, EX4 4QL, UK \label{exeter}
    \and
    ONERA (Office National d'Etudes et de Recherches Aérospatiales), B.P.72, F-92322 Chatillon, France \label{onera}
    \and
    UK Astronomy Technology Centre, Blackford Hill, Edinburgh EH9 3HJ, United Kingdom \label{ukatc}
    \and
    \'Ecole Normale Sup\'erieure, Lyon, CRAL (UMR CNRS 5574), Universit\'e de Lyon, France \label{enslyon}
    \and
    Center for Space and Habitability, University of Bern, 3012 Bern, Switzerland \label{bern}
    \and
    Lunar and Planetary Laboratory, University of Arizona 1629 E. University Blvd. Tucson, AZ 85721, USA \label{tucson}
}

\date{Received 2 February 2021 / Accepted 26 March 2021}

\abstract
{High-contrast imaging techniques combined with medium-resolution spectroscopy can significantly boost the direct detection of exoplanets by using the spectral information of molecular features found in their spectra. HARMONI, one of the first-light instruments to be mounted on ESO's future extremely large telescope (ELT), will be equipped with a single-conjugated adaptive optics system to reach the diffraction limit of the ELT in the $H$ and $K$ bands, a high-contrast module dedicated to exoplanet imaging, and a medium-resolution (up to $R = 17\,000$) optical and near-infrared integral field spectrograph. Combined together, these systems will provide unprecedented contrast limits at separations between 50 and 400~mas. In this paper, we aim at estimating the capabilities of the HARMONI high-contrast module for the direct detection of young giant exoplanets. We use an end-to-end model of the instrument to simulate high-contrast observations performed with HARMONI, based on realistic observation scenarios and observing conditions. We then analyze these data with the so-called “molecule mapping” technique combined to a matched-filter approach, in order to disentangle companions from the host star and tellurics and increase the signal-to-noise ratio of the planetary signal. We detect planets above 5$\sigma$ at contrasts up to 16~mag and separations down to 75~mas in several spectral configurations of the instrument. We show that molecule mapping allows here to detect companions up to 2.5 magnitudes fainter compared to state-of-the-art classical high-contrast imaging techniques based on angular differential imaging. We also demonstrate that the performance is not strongly affected by the spectral type of the host star, and we show that we reach close sensitivities for the best three quartiles of observing conditions at Armazones, which means that HARMONI could be used in near-critical observations during 60 to 70\% of telescope time at the ELT. Finally, we use simulated planets from population synthesis models to further explore the parameter space that HARMONI and its high-contrast module will open compared to current high-contrast instrumentation.}


\titlerunning{Direct imaging and spectroscopy of exoplanets with the ELT/HARMONI high-contrast module}
\authorrunning{Houllé~et~al.~(2021)}

\maketitle

\section{Introduction}

High-contrast imaging (HCI) is the ideal technique to detect and directly study young planetary systems (<~100~Myr) and their evolution after the early stages of planetary formation. So far, it has allowed the detection of a few dozens of brown dwarfs and giant exoplanets around nearby stars (<~150~pc), and has enabled to place tight constraints on the demographics of giant exoplanets in the outer regions (>~10~au) of planetary systems \citep{Nielsen2019, Vigan2020}. The ability to spatially resolve the companions from their host star allows to directly measure their luminosities and spectra, providing important insights on the formation of these objects. The latest generation of HCI instruments, such as VLT/SPHERE \citep{Beuzit2019}, Gemini/GPI \citep{Macintosh2014} or Subaru/SCExAO \citep{Jovanovic2015}, has pushed further the detection limits thanks to extreme adaptive optics and coronagraphy, but has only added five new detections in total to the list of known companions \citep{Macintosh2015, Konopacky2016, Chauvin2017, Keppler2018, Cheetham2018}. These few new discoveries are mainly due to the scarcity of companions in the regions probed by the technique with the current instruments, but population models tend to indicate than the current limits of HCI are close to more populated regions in terms of separation and mass \citep[e.g.,][]{Mordasini2018}. Extremely large telescopes (ELTs) are therefore needed to go deeper and closer in than the current observations.

Although direct imaging is one of the few techniques to provide spectroscopy of companions, current HCI instruments are only equipped with low-resolution ($R=\lambda/\Delta\lambda=30\textrm{--}100$) spectroscopic capabilities which prevent performing advanced studies such as the chemical characterization of atmospheres or measurements of radial velocities. For this reason, medium to high-resolution spectrographs such as Keck/OSIRIS \citep{Larkin2006}, Keck/NIRSPEC \citep{McLean1998} or VLT/CRIRES \citep{Kaeufl2004} have been used for the characterization of previously discovered companions, allowing the detection of molecules and measurements of C/O ratios in atmospheres \citep{Konopacky2013, Wang2018}, Doppler imaging of a brown dwarf's photosphere \citep{Crossfield2014} or measurements of spin velocities \citep{Snellen2014, Schwarz2016}.

In addition to allowing characterization, higher spectral resolutions can also increase the effectiveness of the detection of exoplanets. \cite{Sparks2002} first proposed the combination of medium- to high-resolution spectroscopy with coronagraphic imaging using integral field spectrographs (IFS) for this purpose. The spatial separation between the starlight scattered in speckles all across the field of view, and the planetary signal strongly concentrated at its location, allows to build a model of the star contribution and subtract it from the spectral cubes. The technique relies heavily on the differences in spectral properties between the planet and the host star: it uses cross-correlation with a spectral template to co-add the planetary spectral lines and reject the telluric and stellar ones, ultimately boosting the planetary signal. \cite{Snellen2015} showed the potential of this technique for the future extremely large telescopes (ELTs), demonstrating that Earth-like planets could potentially be detected in one night of observations with a 39-m telescope in the visible and mid-infrared.

More recently, \cite{Hoeijmakers2018} used this technique on the medium-resolution IFS VLT/SINFONI \citep{Eisenhauer2003} to redetect and characterize $\beta$\,Pic\,b \citep{Lagrange2009}. Even though SINFONI is not equipped with a coronagraph, the technique allowed to mitigate speckles in a more effective way than classical high-contrast imaging techniques based on spectral \citep{Racine1999} or angular \citep{Marois2006} differential imaging (ADI). It allowed to get a clearer detection of the companion and to probe closer regions to the star, showing the high effectiveness of combining medium-resolution spectroscopy with direct imaging. Their so-called "molecule mapping" technique was used successfully again on HR\,8799\,b \citep{Petit2018} and HIP\,65426\,b \citep{Petrus2020}, further demonstrating the potential of medium-resolution IFSs. Finally, \cite{Haffert2019} used a similar technique in the visible with VLT/MUSE to not only detect PDS\,70\,b, but also infer the presence of a second planetary companion (PDS\,70\,c) that could not be detected by ADI as it is embedded in the disk. Contrary to the previous examples, their analysis focused on one intense spectral line (H$\alpha$) and did not require using the cross-correlation with a full spectral template. Two asymmetric atomic jets were found in the same fashion with MUSE in HD~163296 by \cite{Xie2020}.

These studies clearly show the potential of medium- to high-resolution multi-spectral approaches for the detection and characterization of exoplanets. New instruments combining high-contrast imaging and medium- to high-resolution spectroscopy are currently in development. At the VLT, the IFS of SINFONI (SPIFFI) is being upgraded with an additional medium-resolution ($R = 8000$) grating and combined with a new high-contrast imager and adaptive optics module into the upcoming ERIS instrument \citep{Davies2018}. Other projects aim to combine existing facilities by a fiber coupling such as KPIC \citep{Mawet2016}, coupling the NIRC2 camera to NIRSPEC at the Keck II telescope, or HiRISE \citep{Vigan2018, Otten2020} coupling SPHERE to the upgraded CRIRES at the VLT. Finally, on the future ELT of the European Southern Observatory (ESO), first-generation instruments such as METIS \citep{Brandl2016} in the mid-infrared and HARMONI \citep{Thatte2016} in the visible and near-infrared will natively include the combination of medium- to high-resolution spectroscopy with high-contrast imaging. In particular, the medium-resolution (up to $R = 17\,000$) IFS of HARMONI will be equipped with a single-conjugated adaptive optics (SCAO) system to reach the diffraction limit of the ELT in the $H$ and $K$ bands, as well as a high-contrast module dedicated to exoplanet imaging.

In the present work, we estimate the capabilities for direct exoplanet detection of HARMONI, by generating realistic simulations of observations performed with its high-contrast module and analyzing them with the molecule mapping technique. First, we introduce in Sect.~\ref{sec:HARMONI_HC_module} the high-contrast module of HARMONI. The simulation model of the instrument, i.e., simulations of the adaptive optics, coronagraphic images and the injection of fake objects and noises, is described in Sect.~\ref{sec:sim_model}. We then describe the setting of the simulated observing sequence and the astrophysical parameters of the injected objects in Sect.~\ref{sec:sim_astro}. In Sect.~\ref{sec:analysis} we explain how we analyze the simulated data using the molecule mapping framework. We present the results in Sect.~\ref{sec:results}, and we conclude in Sect.~\ref{sec:conclusions}.

\section{HARMONI high-contrast module}
\label{sec:HARMONI_HC_module}

\begin{table*}
    \caption[]{HARMONI-HC configurations providing the smallest inner working angle}
    \label{tab:hc_configs}
    \centering
    \begin{tabular}{lccccccc}
    \hline
    Configuration & $\lambda_{\mathrm{min}}$ & $\lambda_{\mathrm{max}}$ & Spectral & Apodizer & IWA\tablefootmark{a} & OWA\tablefootmark{b} & FPM radius\tablefootmark{c} \\
             & [$\mu$m]                   & [$\mu$m]                   &     resolution       &          & [$\lambda/D$] & [$\lambda/D$] & [mas] \\
    \hline
    HK       & 1.450                    & 2.450                    & 3\,555     & SP2    & 7             & 40            & 95 \\
    \hline
    H        & 1.435                    & 1.815                    & 7\,104     & SP1    & 5             & 12            & 45 \\
    K        & 1.951                    & 2.469                    & 7\,104     & SP1    & 5             & 12            & 70 \\
    \hline
    H-high  & 1.538                    & 1.678                    & 17\,385    & SP1    & 5             & 12            & 45 \\
    K1-high & 2.017                    & 2.201                    & 17\,385    & SP1    & 5             & 12            & 70 \\
    K2-high & 2.199                    & 2.400                    & 17\,385    & SP1    & 5             & 12            & 70 \\
    \hline
    \end{tabular} 
    \tablefoot{\tablefoottext{a}{Inner working angle.} \tablefoottext{b}{Outer working angle.} \tablefoottext{c}{Focal plane mask. Note that the radii are only approximate because the masks are slightly asymmetric to accommodate the residual atmospheric dispersion.}}
\end{table*}

HARMONI is a general-purpose first-light instrument for the ELT, which provides various adaptive optics (AO) modes and an integral field unit (IFU) offering different spectral resolutions and spatial samplings. One of the many science cases for HARMONI is the direct detection and characterization of young giant exoplanets \citep{Thatte2016}. For this specific science case, it has been proposed to implement a dedicated high-contrast (HC) module \citep{Carlotti2018}, which provides HARMONI with the capability to directly image companions and disks located as close as 1\,au around nearby stars.

The HC module works with the SCAO mode of HARMONI and implements different strategies for optimizing the baseline performance of the instrument for high-contrast imaging. Firstly, the HC module increases the signal-to-noise ratio (S/N) of off-axis sources by decreasing the intensity of the diffracted light around the star using apodizers known as shaped pupils \citep{Kasdin2003,Carlotti2011}, i.e., pupil binary amplitude masks. The goal of the apodizers is to maximize the attenuation of the diffraction in a given region of the field of view defined by an inner working angle (IWA) and an outer working angle (OWA).

Secondly, the HC module minimizes the non-common path aberrations (NCPA) between the SCAO system and the science detector by using a dedicated wavefront sensor that measures the wavefront as close as possible to the apodizer. It was decided to use a Zernike wavefront sensor, also known as a ZELDA wavefront sensor \citep{NDiaye2013,Ndiaye2016,Vigan2019}, working at 1.175-$\mu$m wavelength. The wavelength is selected to be as close as possible to the science wavelengths in order to limit the chromatic effects.
    
Thirdly, the module implements partially transmissive focal plane masks (FPM) that enable monitoring the position of the star in real time and avoid saturating the detector. The transmission of the FPMs is expected to be approximately 0.0001 in the final system.

We also highlight an important difference with respect to the baseline described in \cite{Carlotti2018}: the current design now includes an atmospheric dispersion corrector (ADC) designed to be optimal at a fixed zenith angle value of 32.6\degr{} at which dispersion will be entirely compensated. This fixed ADC has been added to the HC module to increase the efficiency of the ZELDA wavefront sensor and to decrease the size of the focal plane masks. Some changes have also been made in the specifications of the apodizers.

Finally, the IFU used for the HC observations is the one from HARMONI and therefore offers the same spectral configurations as for the other modes of the instrument. However, due to the requirement of high image quality, the HC module is designed to work only in $H$ and $K$ bands, which limits the number of grism configurations that can be used for the observations. A $J$-band mode is currently being considered with slightly degraded performance, but it is not considered in our present analysis. The different configurations and associated apodizers being studied in this paper are summarized in Table~\ref{tab:hc_configs}. In the \texttt{HK} configuration, only the SP2 apodizer can be used due to the bandwidth. In the other configurations, we choose in our analysis to use SP1 because it provides the smallest possible IWA. However, in practice, it would be possible to use the SP2 apodizer in all configurations.

\section{End-to-end simulation model}
\label{sec:sim_model}

To estimate the performance of the HARMONI-HC module for exoplanet detection, we use an end-to-end simulation model of the instrument. This model enables to simulate sequences of high-contrast images obtained with realistic observing conditions and photometry. The simulation is composed of three distinct parts described in the following subsections: the observing conditions and adaptive optics correction (Sect.~\ref{sec:sim_ao}), the high-contrast images (Sect.~\ref{sec:sim_hc}) and the photometry (Sect.~\ref{sec:sim_phot}).

\subsection{Adaptive optics simulations}
\label{sec:sim_ao}

First of all, we simulate closed-loop adaptive optics sequences to create realistic SCAO wavefront error residual maps  \citep{Schwartz2020} under different seeing conditions (JQ1, JQ2, and JQ3, detailed after). These simulations only include atmospheric effects: telescope and instrument perturbations are added separately (see Sec.~\ref{sec:sim_hc}) based on the error budget planned for the SCAO of HARMONI. Considering that the final performance of high-contrast imaging instruments depends highly on the frame-to-frame point-spread function (PSF) stability, we simulate for each seeing condition a long observing sequence sampled every 1\,min with short closed-loop sequences (typically 120 sequences for a 2~h-long observing sequence). Each of these closed-loop sequence represents a typical HC exposure, with atmospheric parameters (seeing, wind parameters) fixed within the sequence but varying from one sequence to the next. The mean power spectral density of the residual optical path difference (OPD) maps is computed for each HC exposure and fed to the high-contrast image simulator described in Sec.~\ref{sec:sim_hc}. 

We use the Object-Oriented Matlab Adaptive Optics environment \citep[OOMAO;][]{Conan2014} to perform all the SCAO simulations. OOMAO allows simulations with tunable levels of complexity for 1/ a turbulent atmosphere, 2/ a natural guide star, 3/ a telescope pupil geometry, 4/ a deformable mirror, 5/ a wavefront sensor, and 6/ a control loop.

\begin{figure*}
    \centering
    \includegraphics[width=1.0\textwidth]{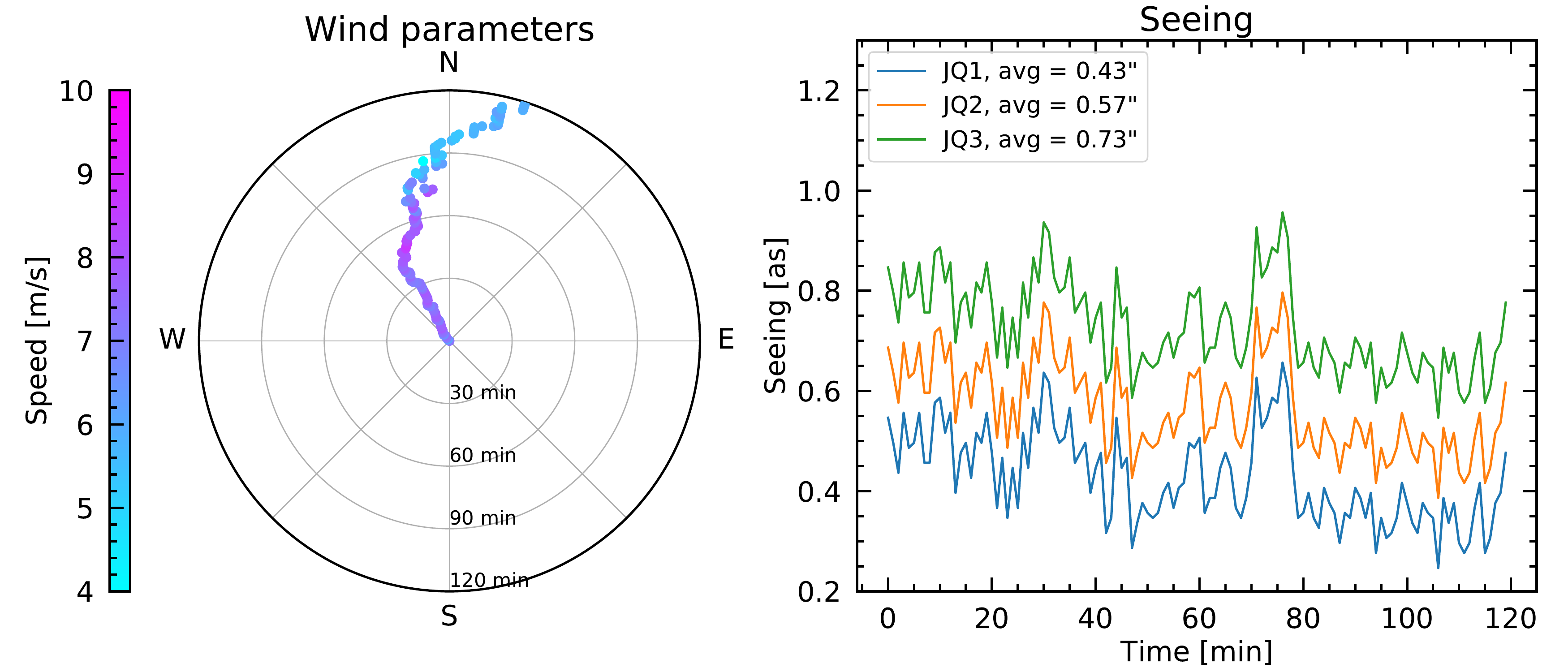}
    \caption{Atmospheric conditions simulated for each exposure. The left plot provides the ground layer wind speed and direction and the right plot shows the seeing values for three different predefined conditions, which are labeled JQ1, JQ2, and JQ3 (see text for details).}
    \label{fig:observing_conditions}
\end{figure*}

For the atmospheric turbulence, we simulate two layers of turbulence: a jet-stream layer at 10~km altitude representing 20\% of the turbulence with fixed wind parameters, and a ground layer at 2.5~km altitude representing 80\% of the turbulence with wind parameters varying between each HC exposure. For the top layer, we simulate a wind speed of 21.2~m/s with a direction -23\degr{} away from North, corresponding to the atmospheric model at Cerro Armazones averaged between 5 and 15~km. To simulate wind variations in the ground layer, we use typical wind measurements from the Cerro Paranal VLT site\footnote{\url{http://archive.eso.org/cms/eso-data/ambient-conditions/paranal-ambient-query-forms.html}}, assuming that the variations will be comparable for the ELT at Cerro Armazones. We use a MASS-DIMM wind speed and direction sequence from a typical observing night at the VLT (MJD 58485) with 120 measurements sampled every minute, and we set its mean value to the wind speed and direction of the atmospheric model at Cerro Armazones averaged between 0 and 5~km, namely $6.4$~m/s and $-10$\degr. The wind sequence simulated for the ground layer is shown in Fig.~\ref{fig:observing_conditions}. To simulate seeing variations between the HC exposures, we similarly use a typical MASS-DIMM seeing sequence (MJD 58488) and we set its mean value to three different levels: 0.43\arcsec\,(JQ1), 0.57\arcsec\,(JQ2), and 0.73\arcsec\,(JQ3). These values correspond to the three most favorable quartiles of seeing level at Cerro Armazones. The fourth quartile is not considered in the current simulations since it is deemed unlikely that the HC module will be used in poor observing conditions. The simulated seeing sequences are shown in Fig.~\ref{fig:observing_conditions} for the three levels. The atmospheric turbulence is simulated using a von Karman distribution with an outer scale of 50~m.

For the natural guide star (NGS), we simulate a source of magnitude 8 in the wavefront sensor bandpass. We simulate a total throughput of 10\% for the atmosphere, telescope, and instrument combined. We change the zenithal angle of the NGS for each of the 120 simulated HC exposure, using the elevation of the star HIP~65426 \citep[known to have a planetary mass companion,][]{Chauvin2017} observed from Cerro Armazones during a 2~h-long observation centered on its transit to meridian at 28.2\degr.

For the telescope, we simulate a 38.542~m-diameter ELT pupil, with the hexagonal footprint of the outer segments partially vignetted by the M3 mirror to a 10\arcmin{} field of view, and the innermost segments fully vignetted up to an 11.208~m-diameter obstruction. The six spider struts are simulated with a width of 50~cm.

For the M4 deformable mirror (DM), we simulate 4672 actuators with a pitch of 50~cm projected in M1 space and Gaussian influence functions with a coupling coefficient of $40\%$. We simulate a DM perfectly aligned with the M1 pupil. To minimize the ``island effect'' caused by the segmentation of the pupil \citep{Schwartz2018}, we enslave the pairs of actuators located on the edge of the spider shadows by summing the influence function of one actuator to the other's \citep{Schwartz2020}. Furthermore, we use a modal approach to control the deformable mirror: we compute the Karhunen-Loeve (KL) modes of the actuators and keep only the first 4000 modes of this basis in the control loop.

For the wavefront sensor, we simulate a pyramid wavefront sensor sensitive in the I3 spectral bandpass (800~nm effective wavelength, 33~nm bandwidth), as will be used for the SCAO system of HARMONI. Our simulations include photon noise and readout noise with 0.3~e$^{-}$/s rms, and a modulation of 3$\lambda/D$ at the tip of the pyramid. The interaction and control matrices linking the wavefront sensor slopes and the DM modes are computed by simulating a bright 0-magnitude calibration source.

Finally, the control loop is simulated with a loop frequency of 500~Hz, a control gain of 0.5, and a 2-frame delay between the integration and the application of the DM command. For each of the 120 HC exposures, we simulate 2550 iterations, which account for 50 iterations necessary for the loop to converge that are discarded to compute the performance (see below). We thus simulate 5~s-long HC exposures, limited by the computation time of the simulations (about 5 days per simulation of 120 HC exposures).

\begin{figure}
    \centering
    \includegraphics[width=\linewidth]{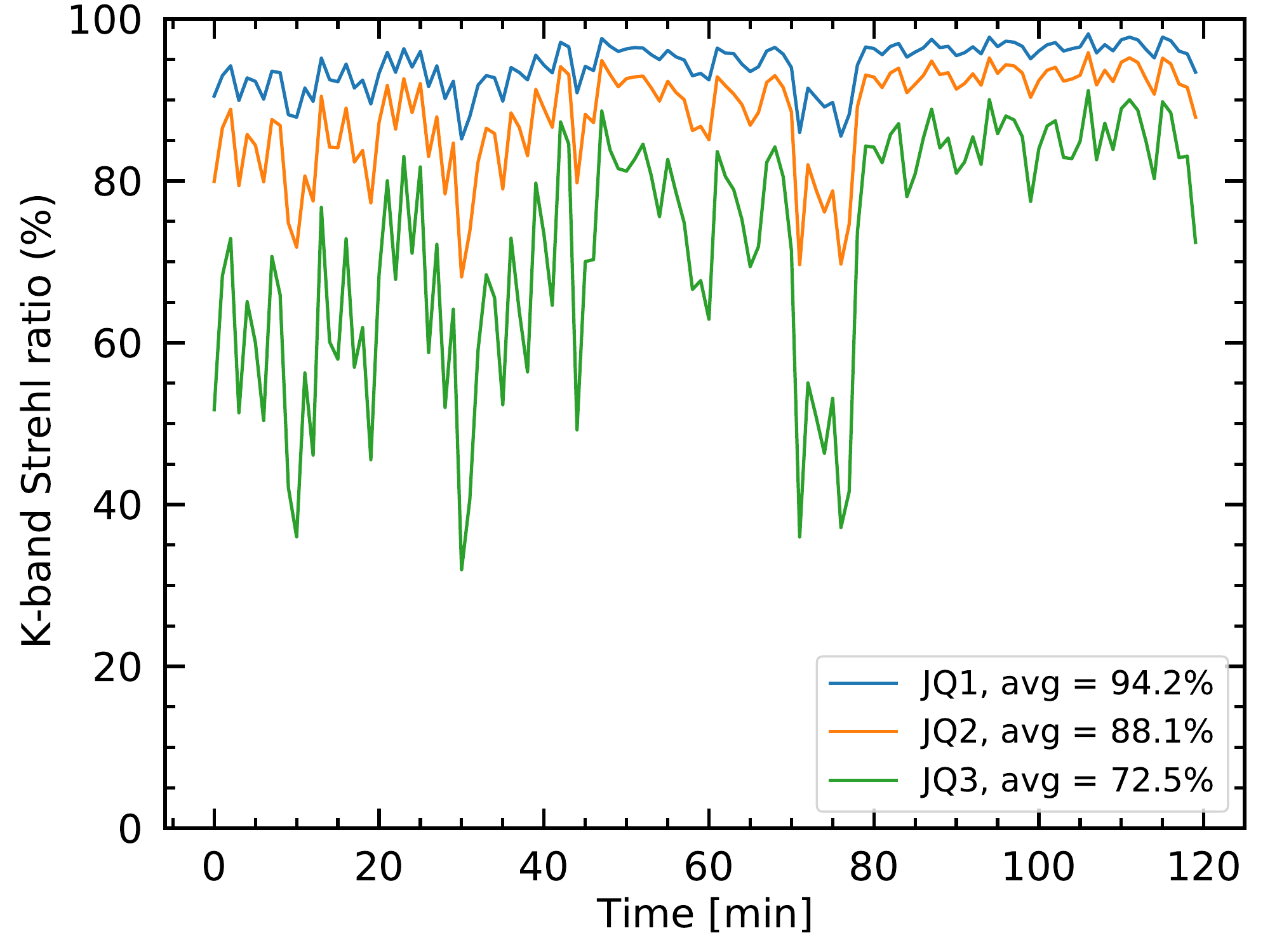}
    \caption{Strehl ratio of the SCAO simulations averaged over the 2500 iterations of each HC exposure, for the three simulated seeing levels. Only atmospheric turbulence is simulated in the closed-loop sequences, additional telescope and instrument effects are separately added to the simulations.}
    \label{fig:Strehl_ratio}
\end{figure}

\begin{table}
    \caption{Summary of HARMONI SCAO performance against atmospheric conditions}
    \label{table:strehl}
    \centering
    \begin{tabular}{lrr}
    \hline
    Seeing level & Average seeing & $K$-band Strehl ratio \\
                 & [\arcsec]          & [\%] \\
    \hline
    JQ1          &     0.43       & 95.8 \\
    JQ2          &     0.57       & 91.4 \\
    JQ3          &     0.73       & 80.3 \\
    \hline \
    \end{tabular} 
\end{table}

From the residual wavefront error of each 2500$\times$120 loop iteration, we compute the Strehl ratio at 2.2~$\mu$m and we average over the 2500 iterations to obtain a Strehl ratio value for each of the 120 long exposures (see Fig.~\ref{fig:Strehl_ratio}). We give the values average over the whole 120 exposures in Table~\ref{table:strehl}. These results reflect only the performance of the SCAO system against atmospheric turbulence over long observing sequences: additional perturbation sources related to telescope and instrument behaviors, such as wind shake and non-common path aberrations (NCPA), are simulated separately (see Sec.~\ref{sec:sim_hc}). The current simulation does not yet include the so-called ``low-wind effect'' \citep{Sauvage2016}. Finally, we compute the power spectral density (PSD) of each HC exposure by averaging the 2500 PSDs of each residual OPD map. These averaged PSDs are then used in the high-contrast simulation module described in the next section.

\subsection{High-contrast images}
\label{sec:sim_hc}

High-contrast images are computed wavelength by wavelength and exposure by exposure through an end-to-end model that takes into account the dispersion of the atmosphere, the key optical components of the system (the telescope aperture, the apodizers, and the focal plane masks), various sources of aberrations (amplitude and phase errors from the telescope, phase errors from the instrument up to the focal plane masks, SCAO residual errors, pupil alignment error), and the ability of the system to sense these aberrations (using the SCAO subsystem and the dedicated ZELDA wavefront sensor), and to correct for them.

Amplitude aberrations are introduced by the primary mirror. Our model includes seven randomly selected missing segments, as well as a reflectivity error randomly chosen for each of the 798 segments and uniformly distributed over a 0--0.05 range. This range corresponds to the typical errors in reflectivity expected for the segments. The primary mirror of the telescope also introduces a phase aberration pattern which is due to cophasing errors. We use as input in the simulation an OPD error map provided by ESO for the primary mirror.

Each of the optics of HARMONI introduce phase aberrations. Our model assumes a $f^{-2}$ power law for these aberrations, where $f$ is the spatial frequency. Their amplitude is derived from an independent error budget analysis \citep{Carlotti2018}. Only optics located upstream of the focal plane masks are taken into account as the masks will block almost all of the stellar light, so the contribution of the downstream aberrations to the scattered light is considered negligible \cite[see, e.g.,][]{Cavarroc2006}.

The motion of the telescope pupil over time is modeled as a smooth cosine function with a 1-h period, and an amplitude that equals 1.6\% of the pupil diameter. This motion translates into a broadband shift of the beam's footprint over the optics, and therefore of the phase aberrations. The system senses part of these phase aberrations, and partially corrects them. Because of atmospheric dispersion, the wavefront measured by the SCAO subsystem at $\sim$0.8~$\mu$m is slightly different from the wavefront errors that induce the aberrations in the science images in the $H$ and $K$ bands. Our end-to-end model takes this chromatic beamshift into account through a ray optics model. As part of our error budget analysis, this ray optics model has been compared to a diffractive model, showing that it was slightly optimistic. The amplitude of the wavefront aberrations used in our end-to-end model reflects the result of this comparison by considering 15\% higher values. Non-common path aberrations are by definition not sensed. They are included in our model, however, and our error budget considers them too.

The ZELDA wavefront sensor is also used to measure the motion of the pupil with respect to the apodizers. Our model assumes that the apodizers' position is corrected with a residual error of 0.2\%. As the apodizers themselves are designed to be robust to a 0.25\% pupil misalignment, this has no impact on the contrast.

The images are computed as the square modulus of the Fourier transform of the electric field derived from our aberration model. To speed up the computation of the images with the residual atmospheric turbulence, we decompose the computation of the PSFs in two parts: first we compute a PSF without the SCAO residuals, and then we use the PSD maps described in Sect.~\ref{sec:sim_ao} to compute an optical transfer function (OTF) of the corrected atmosphere. Taking the Fourier transform of the first image, we compute the OTF of the system independently of the atmosphere. The product of the two OTFs returns the OTF of the system in the presence of SCAO residuals, and the Fourier transform of this expression is the long exposure image. In the end, we take the product of this second image with the focal plane mask, which provides the final focal plane image.

Finally, we simulate the effect of residual wind shake\footnote{The residual wind shake is the wind-induced jitter of the PSF partially corrected by the AO system.} by convolving the image with a 2D Gaussian kernel. Its full width at half maximum in $x$ and $y$ is chosen to match a residual jitter value that is provided by ESO, and we choose to index this value to the seeing regime, so that it has a lower value in JQ1 than in JQ2.

\subsection{Photometry}
\label{sec:sim_phot}

The last part of the simulation takes as input the high-contrast images simulated as described in the previous section. The goal of the photometry module is to apply realistic photometry to the science images, inject fake planets into the data and finally apply the various noise sources induced by the detection process. This module relies heavily on the HSIM tool v300\footnote{\url{https://github.com/HARMONI-ELT/HSIM}} \citep{Zieleniewski2015}, which has been developed as a generic simulation tool for the HARMONI instrument. For technical reasons, HSIM is however not yet compatible with the HARMONI-HC module, so our high-contrast simulations do not use directly HSIM but instead use the same inputs as HSIM (transmission files, thermal background estimates, detection noises, etc).

The input models for the stellar photometry are the BT-NextGen PHOENIX models \citep{Allard2012}. The effective temperature \Teff and surface gravity \logg of the model are chosen to match the stellar spectral type provided by the user. The model is first scaled in amplitude to a specific $H$- or $K$-band magnitude (with respect to Vega) also provided by the user. It is then convolved by a rectangular window with a spectral width corresponding to the size of the spectral resolution elements of the simulated setup. Finally, the convolved model is interpolated at the final wavelengths to obtain the stellar photometry. The user also has the possibility to specify a radial velocity (RV) for the star, in which case the stellar spectrum is Doppler-shifted before the convolution process.

The input models for the planets photometry are the ATMO models (\citealt{Tremblin2017}; \citealt{Phillips2020}). The procedure to obtain the final planetary photometry is similar to the stellar one, except that we scale the spectrum in amplitude by an additional $\Delta$mag provided by the user and corresponding to the contrast between the star and the planets in $H$ or $K$ band. The user can also provide RV and rotational velocity values for the planets, in which case the planet spectrum is Doppler-shifted and rotationally broadened before being convolved.

We use the ESO SkyCalc tool\footnote{\url{https://www.eso.org/observing/etc/bin/gen/form?INS.MODE=swspectr+INS.NAME=SKYCALC}} \citep{Noll2012,Jones2013} to simulate the Earth's atmosphere absorption and emission at high resolution. The simulation is performed for the location of the ELT (Armazones, altitude~=~3060~m) and for each individual airmass value in the observing sequence. This enables to include a realistic variability in the depth of the absorption lines as well as the OH emission lines.

The telescope and HARMONI instrument contributions in terms of absorption and thermal emission are directly based on the HSIM tool. The telescope has a collecting area of 980~m$^2$ and the outside temperature is expected to be at 280~K. The warm part of HARMONI is assumed to be cooled by 20~K with respect to the outside, and the cryostat is cooled down to 130~K. The transmission budget is based on numbers provided by ESO for the telescope ($\sim$75\% in $H$ band) and by the preliminary design study for the HARMONI instrument ($\sim$50\% in $H$ band, excluding the HC module). The transmission of the HC module is estimated to be around 50\% for the shaped-pupil mask and 70\% for all the optics \citep{Carlotti2018}. Finally, the quantum efficiency of the Hawaii-4RG detectors is taken to be $\sim$95\%. Overall, the average transmission in $H$ band reaches $\sim$10\%.

After the photometry is applied to the simulated images at each time step of the simulation, we add the contribution from the sky and thermal background to the images, and we add photon noise following a Poisson distribution. Then we add detector cross-talk (2\% along the spectral dimension and the $x$ spatial dimension), dark current (0.0053~e$^-$/pix/s), and readout noise (12~e$^-$/read). We compute a pseudo-calibration of the sky and thermal background with a different realization of the noise and subtract them from the science images to simulate a realistic calibration process. Finally, we simulate the pipeline interpolation effects by convolving the images with a Gaussian of standard deviation $\sigma = 1$~pix.

\section{Astrophysical simulations}
\label{sec:sim_astro}

In this section we present the general observational and astrophysical assumptions that we use as inputs for the simulations. We treat first the observing sequence and instrumental setups (Sect.~\ref{sec:sim_obs_seq}), and then the astrophysical parameters used for the simulations (Sect.~\ref{sec:astro_params}).

\subsection{Observing sequence}
\label{sec:sim_obs_seq}

The AO simulations and the high-contrast images generation being computationally expensive (several days of computation on a 24-CPU workstation), these simulations cannot be run over a large range of input parameters. This is why we define a realistic baseline observing sequence with predefined observing conditions, stellar declination ($\delta = -15$\degr), and time steps. This sequence is defined to be representative of foreseen observations with the high-contrast mode of HARMONI.

High-contrast imaging observations are typically obtained with the pupil stabilized with respect to the instrument, in order to minimize the rotation of optics and therefore the variation of quasi-static speckles in the focal plane coronagraphic images. This observing strategy, called angular differential imaging \citep{Marois2006}, is implemented as default in all recent exoplanet imaging instruments \citep[e.g.,][]{Beuzit2019,Macintosh2014,Jovanovic2015} and will be in use for the HARMONI-HC module. The main drawback of this observing strategy is that it generally requires to observe the science targets around meridian passage to benefit from the largest amount of field-of-view rotation. For our baseline sequence we adopt a 2-h sequence slightly offset with respect to meridian passage (hour angle from $-0.8$~h to $+1.2$~h) and composed of 120 individual exposures of 60~s each. For a star at declination $\delta = -15$\degr, this results in a total field-of-view rotation of 105\degr.

The observing conditions are classified by ESO into four distinct categories ranging from JQ1 (best quartile of seeing values) to JQ4 (worst quartile of seeing values). By definition, the median conditions are located in between JQ2 and JQ3. For the baseline simulation, we use the JQ2 observing conditions, which correspond to good conditions but not the very best ones expected at the site of the ELT. 

We then compute high-contrast images for the sequence in the six default spectral configurations defined for the HARMONI-HC module (Table~\ref{tab:hc_configs}). Each configuration corresponds to a given spectral domain and resolution. For a given configuration, the optimal shaped-pupil apodizer and focal plane masks are set following a trade-off optimization done in the design study of the HC module \citep{Carlotti2018}.

\subsection{Astrophysical parameters}
\label{sec:astro_params}

\begin{table*}
    \caption[]{Astrophysical parameters of the reference cases}
    \label{tab:sim_params}
    \centering
    \begin{tabular}{lccccccc}
    \hline
    \, & Host star & Companion 1 & Companion 2 \\
    \hline
    Spectral type                 & F0           & T     & L \\
    Effective temperature (\Teff) & 7200 K       & 800 K & 1500 K \\
    Surface gravity (\logg)       & 4.0          & 4.0   & 4.0\\
    Mass                          & 1.45 $M_\sun$ & - & - \\
    Distance                      & 30 pc & - & - \\
    Angular separation           &      -        & \multicolumn{2}{c}{\text{50--275 mas}} \\
    H/K magnitude                 & 4.77         & \multicolumn{2}{c}{\text{4.77 + [9--18 mag]}} \\
    Radial velocity ($v\sin i$)   & 12.6 km/s    & \multicolumn{2}{c}{\text{12.6 + 8 km/s}}\\
    Rotational velocity           &     -         & \multicolumn{2}{c}{\text{20 km/s}}\\
    \hline
    \end{tabular}
\end{table*}

\begin{figure}
    \centering
    \includegraphics[width=\columnwidth]{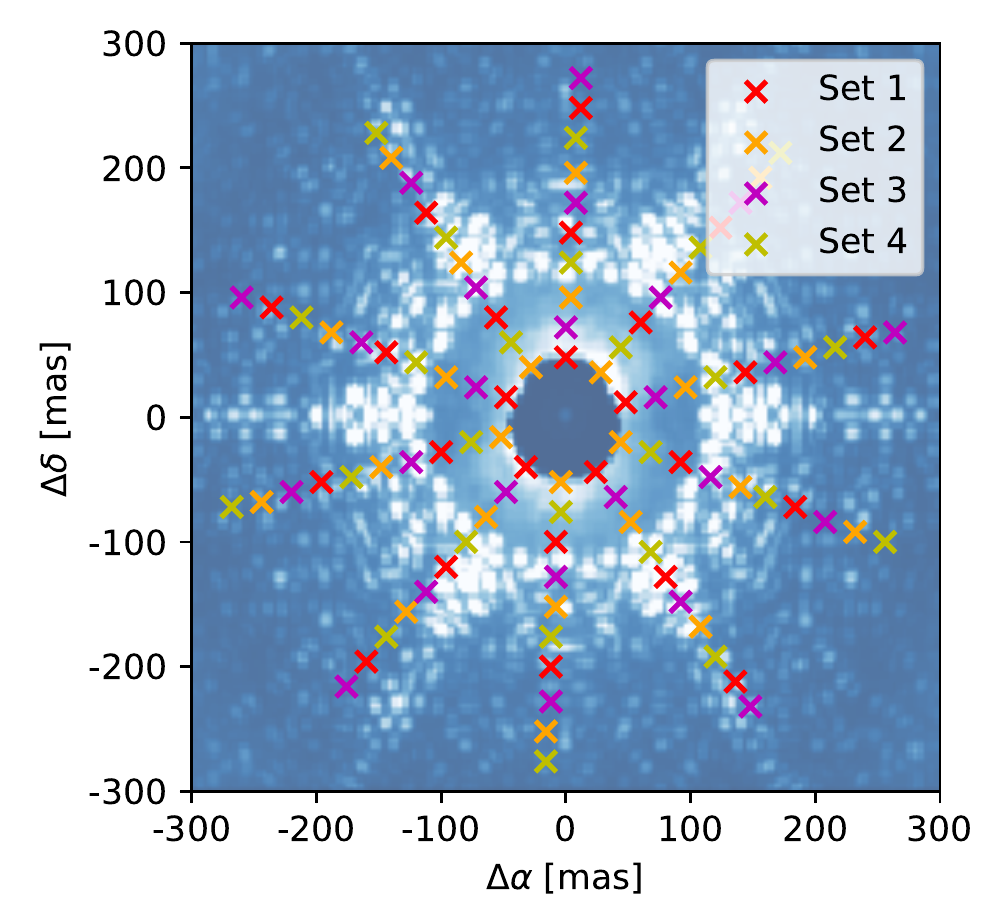}
    \caption{View of the high-contrast images in the \texttt{H} configuration, with the position of injection of the companions overlaid. The companions are splitted in four different datasets to avoid the superposition of their PSFs. The dark asymmetric zone in the center is the focal plane mask, whereas the larger circular one is the zone optimized by the SP1 apodizer (see \citealt{Carlotti2018} for details).}
    \label{fig:injection_companions}
\end{figure}

In terms of astrophysical parameters, we define two reference cases, in which the baseline is the 51~Eri star for the stellar parameters (Table \ref{tab:sim_params}). The two cases differ by the spectral types of the injected companions, for which we choose the T and L spectral types, as they are representative of the objects detected in high-contrast imaging and show interesting spectral differences. A sharp change happens at the transition between L and T, the L-type objects being known to get redder as their temperature decreases, while the early T-type objects get bluer (see reviews by \citealt{Kirkpatrick2005} and \citealt{Baraffe2014}). The first reference case uses the spectrum of a typical T-type companion at $T_{\mathrm{eff}}=800$~K and $\log g=4.0$, close to the parameters of 51~Eri~b (\citealt{Macintosh2015, Rajan2017, Samland2017}), whereas the second case uses the spectrum of a typical L-type companion at $T_{\mathrm{eff}}=1500$~K and $\log g=4.0$, similar to $\beta$~Pic~b (\citealt{Bonnefoy2013, Chilcote2017}).

The field of view of the HC module translates into semi-major axes of 1.35 to 9~au at the distance of 51~Eri (30~pc). In order to choose a typical radial velocity (RV) for the companions, we set the semi-major axis at a middle ground of 5~au, which gives an orbital velocity $v=16$~km/s when combined with the mass of 51~Eri. Taking into account that the probability distribution of sine inclinations ($\sin i$) of stellar systems is uniform between 0 and 1, we take an average $\sin i$ of 0.5, and therefore adopt as a typical companion RV the value $v\sin i=8$~km/s. For the rotational velocity, we choose a conservative value of 20~km/s considering the measurements already obtained for $\beta$~Pic~b \citep{Snellen2014} and GQ~Lup~b \citep{Schwarz2016}, as well the compilations of \cite{Bryan2018, Bryan2020}. These studies show that 20~km/s is a higher limit for young giant companions. We shift and broaden the spectra of the injected planets according to these velocities.

We inject companions at separations between 50 and 300~mas by increment of 25~mas. At each separation, ten planets are injected at evenly spaced position angles of 36\degr. To avoid superposition, this set of 110 companions is split into four complementary datasets independently created and analyzed, with separation increments of 50~mas and position angle increments of 72\degr, as presented in Fig.~\ref{fig:injection_companions}. We generate these four datasets for ten different star-to-planet contrasts $\Delta H$ or $\Delta K$ (depending on the spectral configuration) going from 9 to 18~mag. We simulate this for each one of the six spectral configurations of the HARMONI-HC module described in Table~\ref{tab:hc_configs}, and for the two types of companions. This amounts to a total of 480 different simulations for the reference cases. The results will be presented in Sect.~\ref{sec:results}.

\section{Analysis of the data}
\label{sec:analysis}

Recently, \cite{Hoeijmakers2018} demonstrated the potential of using medium-resolution integral field spectroscopy data for the direct detection of exoplanets. They developed the so-called ``molecule mapping'' technique, in which they use the spectral diversity between the planet and the star to disentangle their signals and boost the planet detectability. First, they model the dominant stellar and telluric components and subtract them from each spaxel of the data cube. Then, they perform a cross-correlation on each spaxel of the residual datacube with a spectral template modeling the companion to enhance the planetary signal. This approach relies on the spatial separation between the planetary signal, which is strongly concentrated at the location of the planet's PSF, and the stellar and telluric ones that are scattered across the field of view.

\subsection{Star and tellurics subtraction}
\label{sec:star-subtraction}

We use the procedure described in \cite{Haffert2019} to model and subtract the stellar and telluric signatures from the spectral datacubes. First, we build a reference spectrum by taking the median of all spaxels normalized to the same flux level. This reference spectrum is divided from each spaxel. At this point, spatially varying low-order residuals still remain. We model them using a Savitzsky-Golay filter of order 1 and a window of 101 wavelength steps, and the resulting model is divided from each spaxel. We then use a principal component analysis (PCA) to model and remove the remaining high-order residuals in the flattened spectra, which are dominated by residual telluric lines. We subtract the first 20 modes from the residual cubes. Finally, we derotate the 120 individual exposures according to their associated parallactic angle to align them to a common reference frame, and we average them to produce the final datacube.

\subsection{Cross-correlation with spectral templates}

In order to boost the planetary signals, we cross-correlate each spaxel of the average datacube with a spectral template that models the companions. We use different families of template models for the injection and the cross-correlation in order to simulate slight discrepancies between the observations and the models used for detection. We use ATMO models \citep{Phillips2020} for the injection (see Sect.~\ref{sec:sim_phot}) and BT-Settl models \citep{Allard2013} for the detection, using a template at the same temperature and surface gravity as the one used for the injected planets. This approach is slightly optimistic, since in reality one would have to cross-correlate with a large library of models, but we had to restrict the number of templates due to the large number of simulations and the associated computational time.

Before cross-correlation, we convolve the template to the spectral resolution of the simulated HARMONI configuration and interpolate it along the associated wavelength grid. We then apply a Savitzsky-Golay filter of the same order and window size as the one used on the observations in Sect. \ref{sec:star-subtraction}. Finally, we Doppler-shift the processed template to a grid of radial velocities $v$ between $-500$ and $+500$~km/s by steps of 1~km/s, giving a final two-dimensional $(\lambda, v)$ template matrix.

The optimal signal-to-noise for the cross-correlation between a noisy signal and a template is given by matched filtering. Its use was already described in the context of high-contrast imaging by \cite{Cantalloube2015} and \cite{Ruffio2017}, but only applied in the spatial dimension. Matched filtering allows to directly obtain the S/N of the cross-correlation function (CCF). Without a matched filter, an alternative method to derive the S/N of a simple CCF is to divide the peak value by the standard deviation in the wings of the CCF. However, this technique leads to a saturation of the S/N at low contrasts because the wings of the CCF do not only contain uncorrelated noise but also the auto-correlation signal, which becomes dominant in the high signal-to-noise regime. Matched filtering takes into account the auto-correlation signal and prevents this effect.

Assuming the residual noise is Gaussian, the S/N of the matched-filter CCF at a given radial velocity $v$ is given by:
\begin{equation}
   \mathrm{S/N}_v = \frac{\sum_i s_i t_{v, i} / \sigma^2_i}{\sqrt{\sum_i t_{v, i}^2 / \sigma^2_i}},
   \label{equ:snr_ccf}
\end{equation}
where $i$ is the wavelength, $s$ the observed spectrum, $t$ the spectral template shifted at a given radial velocity $v$, and $\sigma^2$ the variance of the observed spectrum. We note that the sum can be performed on the spatial dimensions as well, if an appropriate template is provided. To estimate the variance from the observations, at least one of the variables (wavelength, position or time) has to be used to compute the variance over, which means that we assume the variance to be almost constant over that variable. We choose to compute the variance over the time axis, which allows to get an estimation of the noise for every single pixel of the average datacube. We compute it by stacking the 120 derotated individual datacubes. The final $(x, y, v)$ S/N map is obtained by evaluating the S/N according to Eq.~\eqref{equ:snr_ccf} at every spaxel of the average datacube, using a spectral template shifted at velocities between $-500$ and $+500$~km/s, as described previously.

At the end, we normalize the S/N map radially following the method of \cite{Cantalloube2015}, also used in \cite{Ruffio2017}. This step is needed as the S/N can be biased by overly optimistic assumptions on the Gaussianity of the noise assumed in Eq.~\eqref{equ:snr_ccf}, which is not fully verified in regions dominated by speckle noise. This effect can be corrected by normalizing radially the S/N map by its own empirical standard deviation in order to get a standard deviation of one all across the map. In order to prevent the companions from biasing the estimation, the noise is computed as the median absolute deviation (MAD) of the S/N in annuli of width $1 \lambda/D \approx 2$~pix, by steps of 1 pixel. We scale this value to obtain the standard deviation ($\sigma = 1.4826$~MAD for Gaussian distributions), which we then divide radially and velocity by velocity from the S/N maps.

\subsection{Generation of detection limits}

\begin{figure*}
    \centering
    \includegraphics[width=\textwidth]{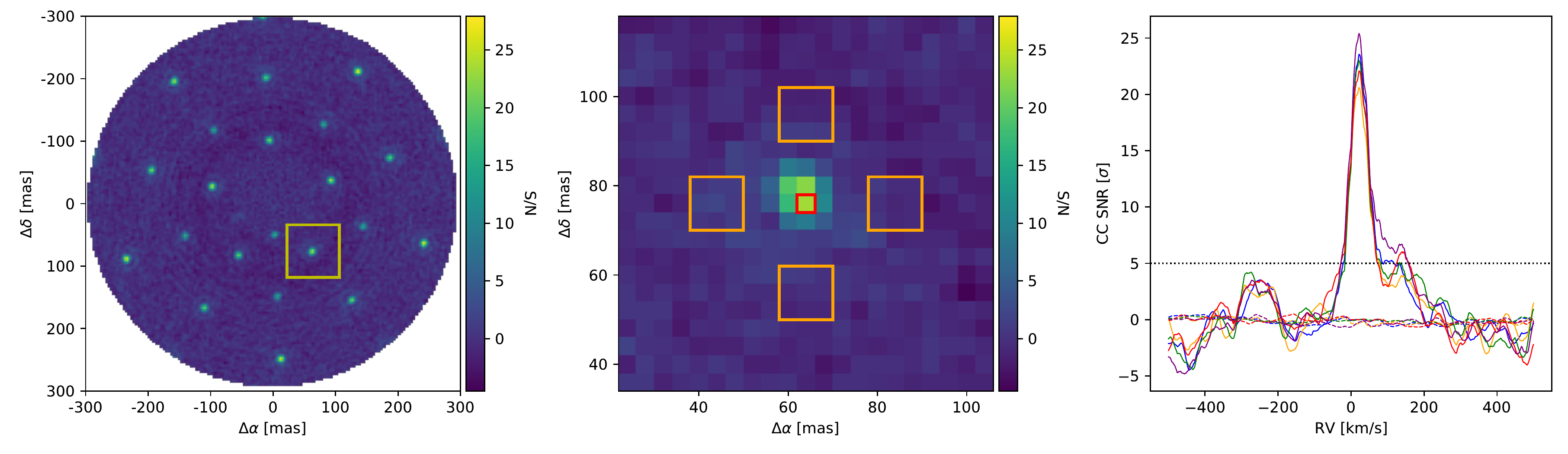}
    \includegraphics[width=\textwidth]{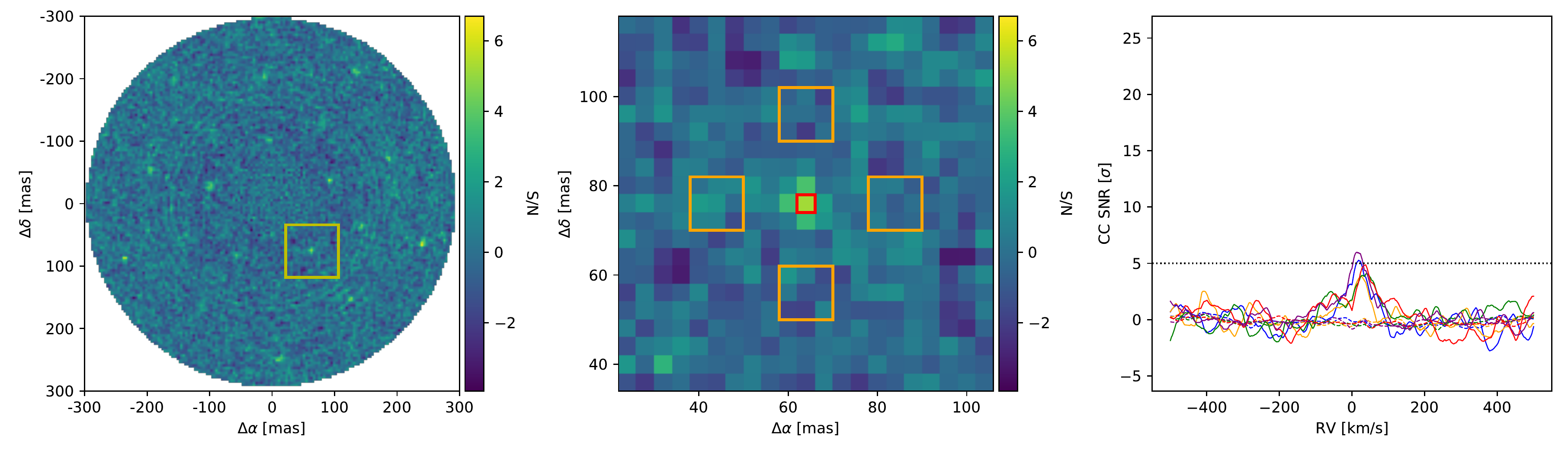}
    \caption{Cross-correlation maps (left) for the \texttt{H-high} configuration with T-type companions at $\Delta H = 14$ (top) and 16~mag (bottom). The central images show enlarged regions around one of the companions at 100~mas, indicated by the yellow squares in the left images. In the right plots we show the corresponding cross-correlation functions (CCF) of this companion and others at same separation, in different colors. The plain lines show the CCFs at the position of injection of the companions (red square in the central image), while the dashed lines show the average CCFs in their neighborhood (taken in the orange squares in the central image). The black dotted line shows the 5-$\sigma$ detection threshold.}
    \label{fig:CCFs-faint-H-high}
\end{figure*}

To estimate the S/N at a given contrast and separation, we extract the S/N values of the companions at their known location of injection and we average them for companions located at the same angular separation (ten samples) and same contrast to produce a final S/N value. We finally interpolate into the grid to compute the 5-$\sigma$ detection limits of the simulated observations.

To compare these results to the performance of classical ADI algorithms currently used in high-contrast imaging, we also run the state-of-the-art ADI algorithm ANDROMEDA \citep{Cantalloube2015} on each configuration. The detection limits from ANDROMEDA are derived from the radial noise standard deviation of the datacubes measured after post-processing, and does not rely on the injection of planets such as for our molecule mapping approach. ANDROMEDA produces S/N maps for each spectral channel and all the channels are combined assuming equal weights.

In order to check the significance of the CCF peaks for the faintest companions, which defines the final sensitivity limits, we plot in Fig.~\ref{fig:CCFs-faint-H-high} the S/N maps and the CCFs of T-type companions at $\Delta H = 14$ and 16~mag in the \texttt{H-high} configuration, both at their location and in their neighborhood. The CCFs of the neighboring pixels shows a flat profile centered around zero and lower than $1\sigma$, which gives confidence in the peaks found on the companions.

\section{Results}
\label{sec:results}

We present in Sect.~\ref{sec:results_ref_cases} the results of the simulations for the reference cases previously described in Sect.~\ref{sec:sim_astro}. Then we show in Sect.~\ref{sec:influence_input_assumptions} the results of additional simulations designed to study the influence of several important input parameters, namely the spectral type and magnitude of the host star, and the observing conditions. Finally, in Sect.~\ref{sec:population_models}, we present results of simulations based on the injection and detection of companions from population models.

From hereon we refer to the \texttt{HK} configuration as low-resolution configuration, to the \texttt{H} and \texttt{K} configurations as medium-resolution configurations, and to the \texttt{H-high}, \texttt{K1-high} and \texttt{K2-high} configurations as high-resolution configurations.

\subsection{Reference cases}
\label{sec:results_ref_cases}

\begin{figure*}
    \centering
    \includegraphics[width=\textwidth]{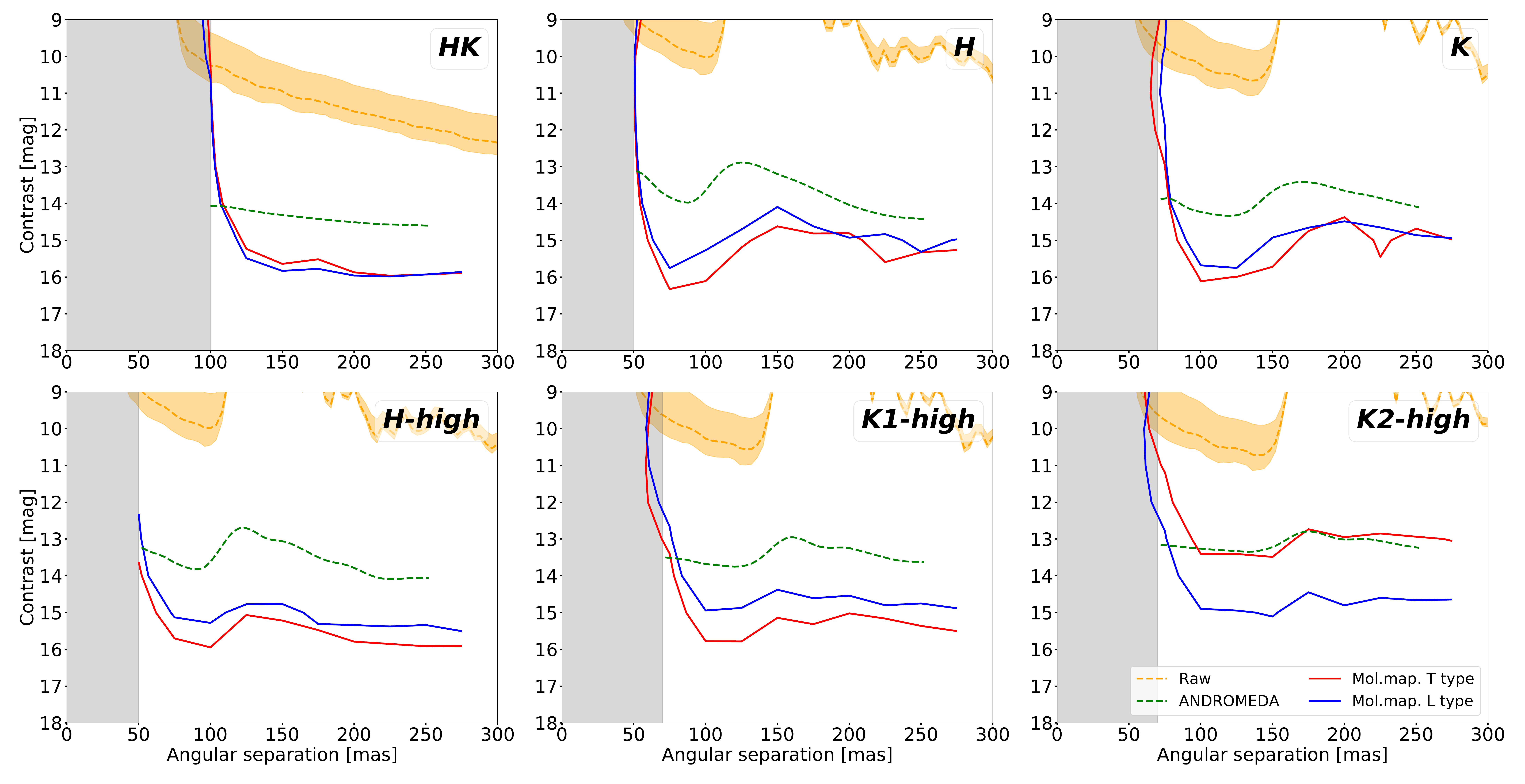}
    \caption{Detection limits of the HARMONI high-contrast mode for the six spectral configurations and the two types of injected companions. The red and blue curves show the 5-$\sigma$ sensitivity to the T-type and L-type companions, respectively. The dashed green curves show the 5-$\sigma$ sensitivity from the processing using the classical angular differential imaging algorithm ANDROMEDA. The orange dashed curves show the raw HARMONI contrasts before any processing, with the envelope representing the 1$\sigma$ variance of the 120 exposures. The grayed-out part corresponds to the area masked by the anti-saturation focal plane mask.}
    \label{fig:SNR-ref-cases}
\end{figure*}

The contrast limits for the reference cases described in Sect.~\ref{sec:sim_astro} are shown in Fig.~\ref{fig:SNR-ref-cases}. We reach a 5-$\sigma$ sensitivity to T-type companions at contrasts up to $\Delta H/K = 16$~mag for all configurations except \texttt{K2-high} at $\Delta K = 13.5$~mag. For the L-type companions, the 5-$\sigma$ sensitivity is reached at $\Delta H = 16$~mag for the low-resolution configuration and 15~mag for the medium- and high-resolution configurations.

The best sensitivities are found between 75 and 100 mas in the $H$-band configurations, between 100 and 125 mas in the $K$-band ones, and above 150 mas in the \texttt{HK} configuration. All medium- and high-resolution configurations show a zone of slightly lower sensitivities between the aforementioned regions and $\sim$200~mas. This zone of lower sensitivities seems to start at the outer working angle of the apodizer (see Table \ref{tab:hc_configs}), i.e., outside of the zone where the raw contrast is optimized by the apodizer. The medium-resolution configurations show a larger variation of contrast limits between the optimized and unoptimized zone, with contrast limits lowered by 1~mag between 100 and 200~mas, whereas the high-resolution ones show lower variations of 0.5~mag. A possible explanation is that the optimized zone experiences a larger broadening throughout the spectral range due to the larger bandwidth of the medium-resolution configurations. As such, at medium resolution, a spaxel in this zone will mix more unoptimized wavelength bins with optimized ones during the cross-correlation operation, which in the end lowers the sensitivity. This effect is less present at larger separations because the scattered starlight decreases, as seen in the raw contrast curves. On the contrary, the low-resolution \texttt{HK} configuration shows almost no radial variation due to its different apodizer, which has an optimized zone almost as large as the field of view but also a larger inner working angle \citep{Carlotti2018}. The effect of the inner working angle set by the focal plane mask is clearly shown with sensitivities strongly decreasing at 100~mas for the \texttt{HK} configuration and 50~mas for the others.

The limits for the L-type companions in configurations other than \texttt{HK} and \texttt{K2-high} are one magnitude worse than for the T-type companions. This can be explained by a lower molecular diversity at higher temperatures for the L types and thus a lower number of spectral lines, which decreases the effectiveness of molecule mapping. The exception of \texttt{K2-high}, which has higher sensitivity to L-types rather than T-types, can be explained by the increased intensity of the CO lines in the L-type objects, and the fact that these lines are a dominating component in this spectral region. The dominance of CO lines over other molecules in the 2.2--2.4~$\mu$m region and their absence in T-type objects would also explain the worse sensitivity to T-type objects in \texttt{K2-high} compared to the other configurations.

The contrast curves of the ANDROMEDA angular differential imaging algorithm show the same profile as the molecule mapping ones, but usually at a worse contrast. The gain of using molecule mapping is increasing with the spectral resolution: the \texttt{HK} configuration shows a gain of 1.5~mag whereas the \texttt{H-high} and \texttt{K1-high} goes up to 2.5~mag. This confirms that the molecule mapping technique efficiently uses the higher amount of spectral lines at higher resolutions to increase the sensitivity to planetary companions. We also note that although the multi-channel S/N maps of ANDROMEDA are combined with equal weights, using a weighting scheme base on L- and T-type spectral templates does not improve the ANDROMEDA sensitivity. The reliance of molecule mapping on the spectral diversity is further shown by the \texttt{K2-high} configuration, where the lower spectral content and the higher thermal background in this band only allows molecule mapping to perform equally as angular differential imaging for the T-type companions.

\subsection{Influence of input assumptions}
\label{sec:influence_input_assumptions}

In this section we analyze the impact of three important input parameters: the host star's spectral type, the host star's magnitude, and the quality of the observing conditions. For these new simulations we focus on the \texttt{HK} and \texttt{H-high} configurations, which are representative of the lowest and highest available resolutions. The T- and L-type companions show no difference in the shape of the detection limits for these simulations, outside of their different levels highlighted in the previous section. For clarity we only show hereafter the detection limits of the T-type companions, but the results are valid for the L-type companions as well.

\subsubsection{Stellar spectral type}
\label{sec:results_stellar_type}

We compare here the results for different host star's spectral types. Stars of the late spectral types such as M and K possess a significant amount of molecular lines that can potentially blend with the planetary ones. It could in principle lower the sensitivity to the companions around host stars of late spectral types, if the stellar component is not well removed before cross-correlation.

\begin{figure*}
    \centering
    \includegraphics[width=\textwidth]{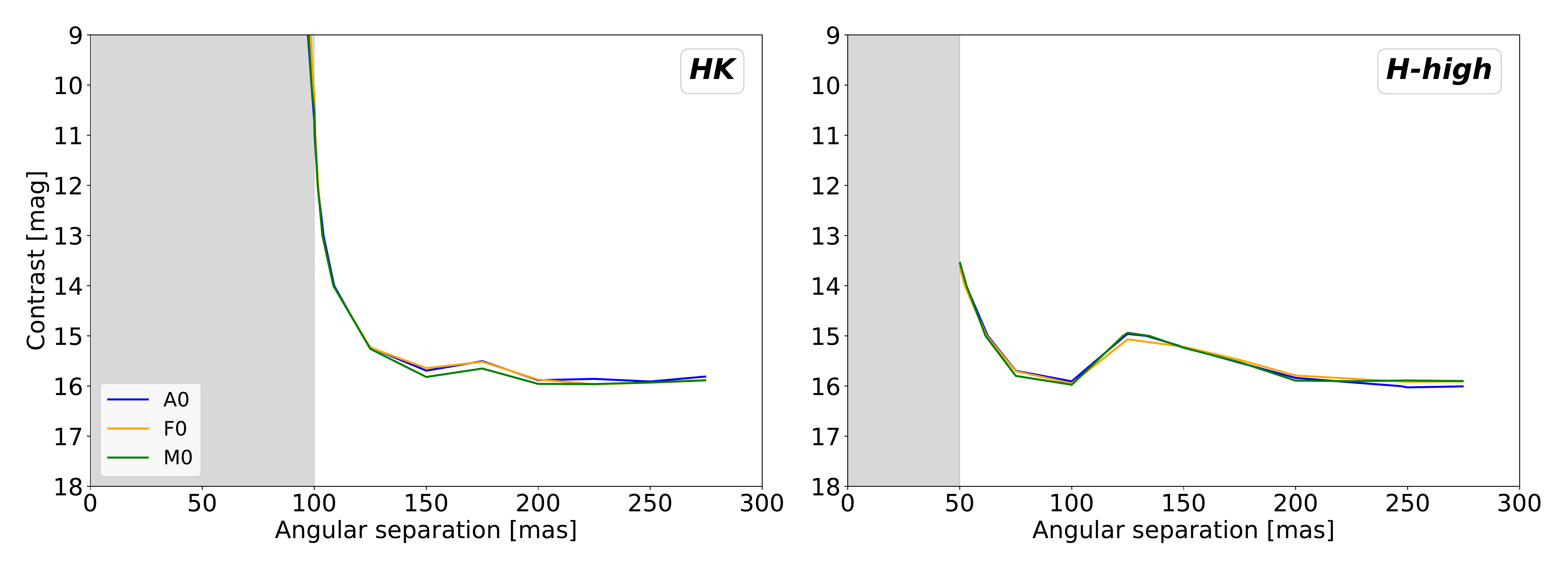}
    \caption{Detection limits of the HARMONI high-contrast mode for the \texttt{HK} and \texttt{H-high} configurations and three different stellar spectral types. The grayed-out part corresponds to the area masked by the anti-saturation focal plane mask.}
    \label{fig:SNR-stellar-types}
\end{figure*}

In Fig.~\ref{fig:SNR-stellar-types} we show the T-type detection limits obtained for A0, F0, and M0 stellar spectral types. These results show no significant differences when the spectral type of the injected host star is modified, even for early M stars. This confirms the efficiency of the stellar subtraction during the process described in Sect.~\ref{sec:analysis}, which leaves stellar residuals at a level weak enough to not influence significantly the sensitivity to the planetary companions. We also see that the behavior is identical between the \texttt{HK} and \texttt{H-high} configurations, which confirms that the stellar spectral type has no significant influence on the final performance.

\subsubsection{Stellar magnitude}
\label{sec:results_stellar_mag}

\begin{figure*}
    \centering
    \includegraphics[width=\textwidth]{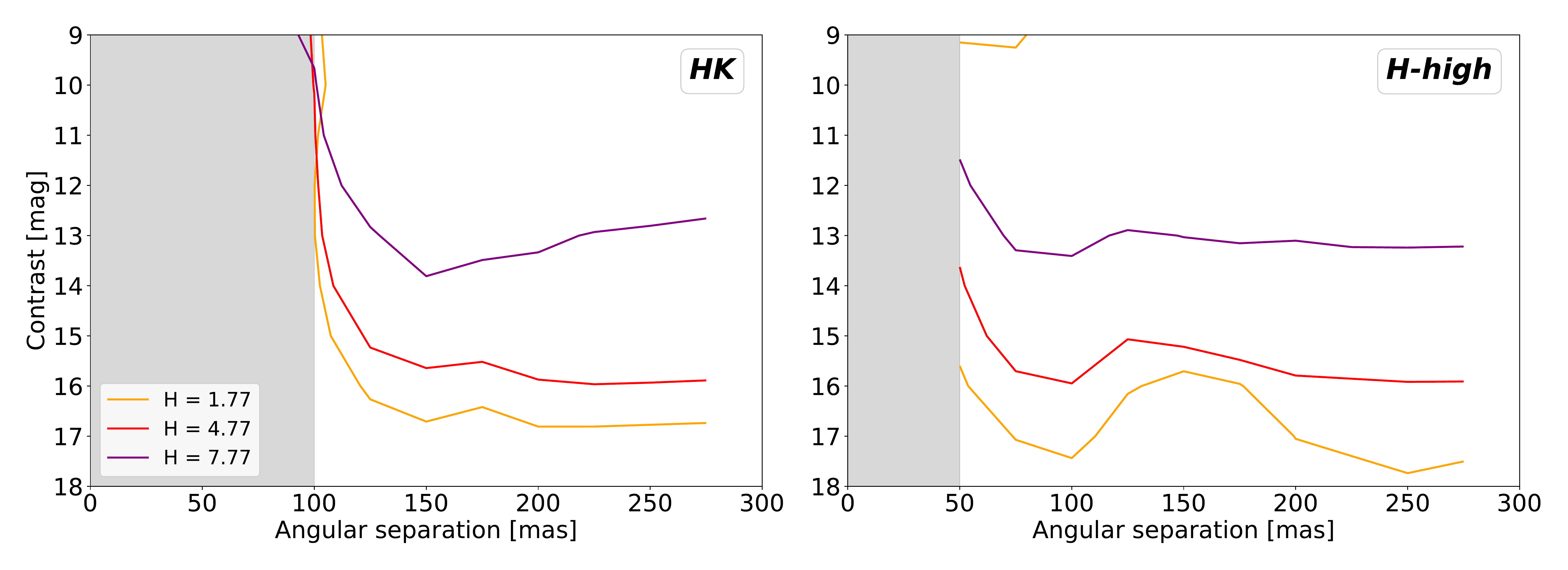}
    \caption{Detection limits of the HARMONI high-contrast mode for the \texttt{HK} and \texttt{H-high} configurations and three different stellar magnitudes. The grayed-out part corresponds to the area masked by the anti-saturation focal plane mask.}
    \label{fig:SNR-dmag}
\end{figure*}

The stellar magnitude is also an important parameter because the luminosity of the companions in the simulations is defined as a contrast to their host star. Varying the magnitude of the host star therefore affects the magnitude of the companion in the same amount. We choose to simulate stellar magnitudes three magnitudes fainter and brighter than the reference cases. Contrast limits for the different stellar magnitudes are shown in Fig.~\ref{fig:SNR-dmag}.

For a very bright host star at $H = 1.77$, we obtain contrast limits improved by 1~mag compared to the reference case. On the other side, much fainter host stars at $H = 7.77$ show contrast limits worse by $\sim$3~mag. The contrast limit does not linearly follow the host star magnitude, which is due to different noise regimes dominating at different magnitudes. The faint star curve does not follow a flat radial profile in \texttt{HK}, indicating that the noise over 150~mas comes mostly from instrumental effects rather than Gaussian photon noise in the lowest signal-to-noise regimes. In the bright star case, the data is dominated by the bright speckles, while in the faint star case the thermal background and readout noise become dominating noise sources.

The reference case at $H = 4.77$ is typical of AF stars at $\sim$30~pc, such as 51~Eri. The simulations show that companions with contrasts up to 16~mag can be detected in 2-hour exposure time around these stars. A planet like 51~Eri~b ($\Delta H = 14.8$) could easily be detected in this amount of time. On the other side, AF stars in nearby OB associations, such as the Scorpius-Centaurus association (\citealt{deGeus1989, deZeeuw1999}), have $H$ magnitudes around 8, similar to the $\Delta H = 7.77$ case. For these stars, detections up to 13~mag are possible, which represents most of the companions already detected so far by high-contrast imaging, but the diameter of the ELT offers here a significant gain in angular resolution.

\subsubsection{Observing conditions}
\label{sec:results_obs_conditions}

\begin{figure*}
    \centering
    \includegraphics[width=\textwidth]{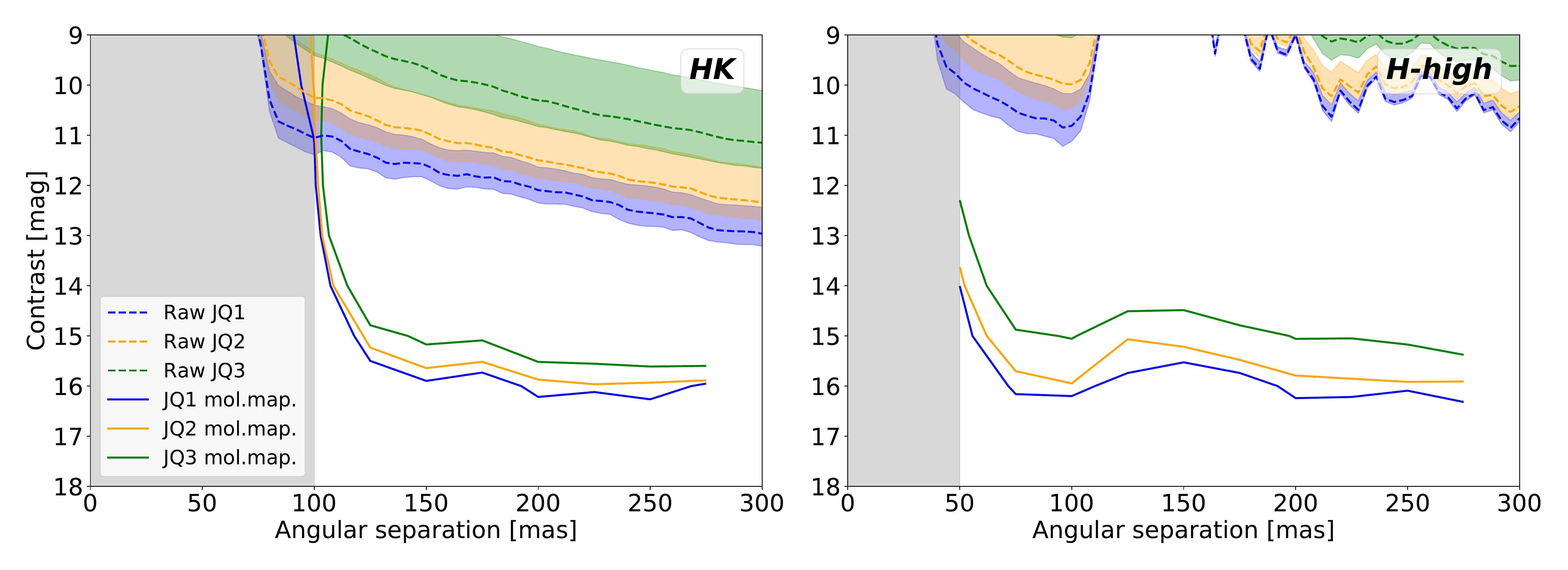}
    \caption{Detection limits of the HARMONI high-contrast mode for the \texttt{HK} and \texttt{H-high} configurations and three different observing conditions. The dashed curves in the upper part of the plots show the raw HARMONI contrasts before any processing, with the envelope representing the 1$\sigma$ variance of the 120 exposures. The grayed-out part corresponds to the area masked by the anti-saturation focal plane mask.
    }
    \label{fig:SNR-JQs}
\end{figure*}

Finally, we look at the expected performance for different seeing conditions. The seeing is a driving parameter for AO systems, so it can affect significantly the performance in HCI observations. We generate simulations for the JQ1 (seeing~=~0.43\arcsec), JQ2 (seeing~=~0.57\arcsec) and JQ3 (seeing~=~0.73\arcsec) conditions defined by ESO, and the contrast limits are shown in Fig.~\ref{fig:SNR-JQs}.

The JQ2 and JQ3 quartiles respectively decrease the sensitivity compared to JQ1 by 0.2 and 0.8~mag in the \texttt{HK} configuration, and by 0.4 and 1~mag in the \texttt{H-high} configuration. The close performances for the JQ1 and JQ2 observing conditions are a very important finding, as it means that optimal performances can be obtained for almost half of the observing time at the ELT. We can further argue that the performance in JQ3 conditions is quite close to JQ1 and JQ2 as well, thus allowing to schedule slightly below critical observations during 60 to 70\% of telescope time at the ELT. These results indicate that the scheduling of exoplanet observations with HARMONI could be relatively easy and not strongly constrained by the environmental conditions.

Varying seeing conditions produce slightly more significant differences on the performance between the \texttt{HK} and the \texttt{H-high} configurations than other input parameters, with the high-resolution configuration being more impacted by the change. One of the main differences in the two configurations is the inner working angle of the apodizer, which is $\sim$50~mas smaller in \texttt{H-high}, and the size of the optimized region, which is much wider in \texttt{HK}. We assume that the observed performance differences are due in part to the differences in the apodizer design that affects the amount of scattered starlight in the focal plane. Further simulations would be needed to confirm this assumption, but our conclusion that the observing conditions have no major impact between JQ1 and JQ2 remains valid.

\subsection{Population models}
\label{sec:population_models}

\begin{figure*}
    \centering
    \includegraphics[width=\textwidth]{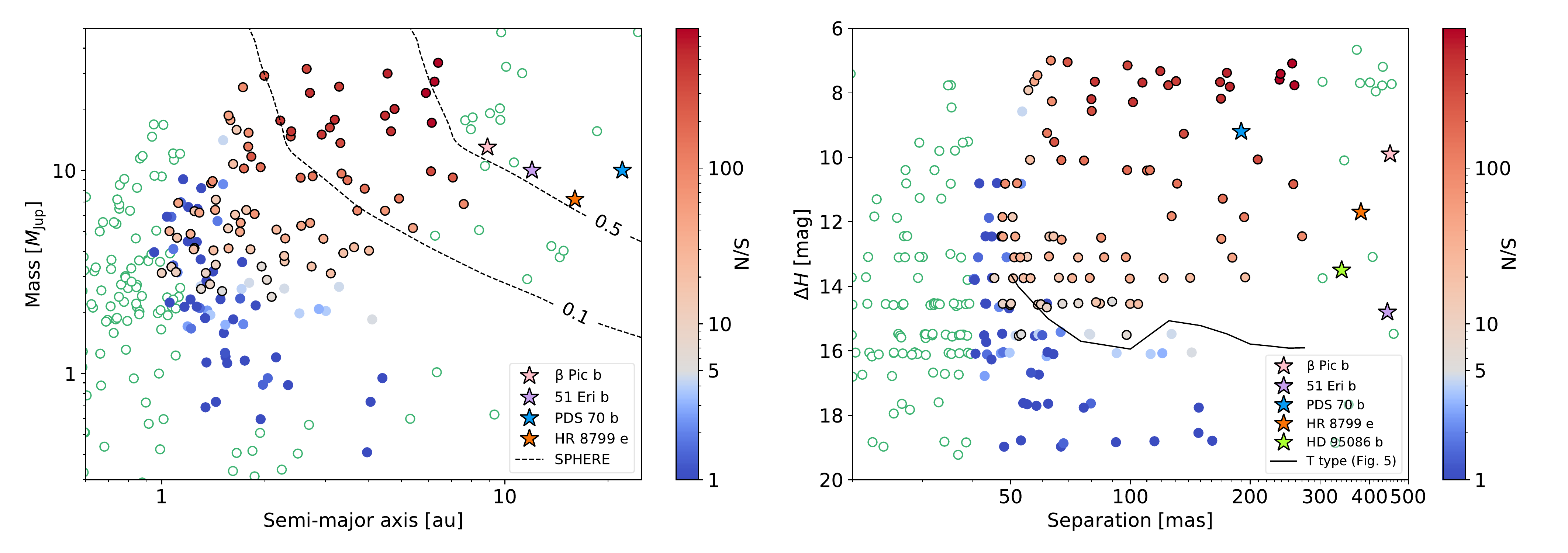}
    \caption{Results of our detection simulation based on population synthesis models, using the \texttt{H-high} configuration, on a semi-major axis/mass space (left) and separation/contrast space (right). Blue to red filled dots are the simulated companions, colored according to their S/N. The transition from blue to red is set at 5$\sigma$ to highlight the detection limit. The companions detected over 5$\sigma$ are further circled in black for clarity, and the green empty dots show the remaining companions of the population that were not simulated (see text for details). Some known companions are included for comparison. We show as well in the left plot the 10\% and 50\% detection probability curves of the VLT/SPHERE SHINE survey from \cite{Vigan2020}. In the right plot we overlay the T-type detection limit in \texttt{H-high} from Fig. \ref{fig:SNR-ref-cases}.}
    \label{fig:synth_pop}
\end{figure*}

The previous simulations provide a clear assessment of the detection capabilities of HARMONI for low-mass planetary companions at various contrasts. However, they are not fully realistic in the sense that the L- and T-type companions have been arbitrarily scaled in contrast in order to cover the full scale of contrasts expected to be accessible to the HC module. A more realistic approach would be to scale the \Teff and \logg of the companions injected into the simulations based on evolutionary tracks \citep[e.g.,][]{Baraffe2003,Marley2007,Spiegel2012,Marleau2019}, or to even use the direct output of population synthesis models, which will not only predict the physical properties of the companions, but also their key orbital parameters, such as the semi-major axis and eccentricity \citep[e.g.,][]{Mordasini2009,Forgan2013}.

In the present work we use the output at 20~Myr of population \texttt{NG76} from the new generation planetary population synthesis (NGPPS) model from Bern \citep{Mordasini2018,Emsenhuber2020a,Emsenhuber2020b}, which is based on the core accretion planet formation paradigm \citep{Mizuno1980,Pollack1996,Alibert2004}. The model provides in particular the mass ($M_p$), radius ($R_p$), and luminosity ($L_p$) of the planet. To infer the appropriate \Teff of the simulated planets, we use the Stefan–Boltzmann law that ties \Teff with $L_p$ and $R_p$ so that:
\begin{equation}
    \label{eq:lum}
    L_p = 4\pi R_p^{2}\sigma {T_\mathrm{eff}^{4}},
\end{equation}
\noindent where $\sigma$ is the Stefan–Boltzmann constant. Similarly, the surface gravity $g$ is computed based on $R_p$:
\begin{equation}
    g = \frac{GM_p}{R_p^2},
\end{equation}
\noindent with $G$ the gravitational constant. Based on the values computed for \Teff and \logg, we then associate to each planet the closest model available in the ATMO spectral library, and we compute the integrated flux and corresponding contrast in $H$ band with respect to our reference star (see properties in Table~\ref{tab:sim_params}). The semi-major axis and eccentricity provided by the population model are converted into angular separation assuming a distance $d = 30$~pc for the system. We assume that each system is observed at maximum elongation and face-on, which translates into the maximum possible angular separation. Finally, we simulate 2-hour observations of each system with the \texttt{H-high} configuration of HARMONI in the same fashion as described in Sect.~\ref{sec:sim_astro} and we analyze the data as described in Sect.~\ref{sec:analysis}. In order to reduce the total number of simulations, we restrict them to planets that fall in the separation range 40--310~mas, which corresponds to the simulated field of view not occulted by the focal plane mask, and in the $H$-band contrast range 6--20~mag, which largely covers the sensitivity range of the high-contrast mode according to Sect.~\ref{sec:results_ref_cases}. As we assume that the planets are observed at maximum elongation, planets with apocenters falling outside the field of view are not simulated, even if their semi-major axis falls inside. The simulated subset translates to $T_{\mathrm{eff}}$ between 400 and 2700~K, and $\log g$ between 2.5 and 4.5.

The detection results are presented in Fig.~\ref{fig:synth_pop}. We find that all companions in the field of view of the instrument with masses $>3~M_{\mathrm{Jup}}$ and semi-major axes $>2$~au are detected at 5$\sigma$ or higher confidence. Companions down to 1~au and $2~M_{\mathrm{Jup}}$ are partly detected as well. A clearer view of what conditions the detectability in this region is offered on the separation--contrast plot, where hard limits of detection can be seen around 50 mas and 16 mag. The reasons some companions between 1 and 2~au are not detected are either that their position angle and separation, computed from the maximum elongation, put them behind the asymmetric focal plane mask (see Table \ref{tab:hc_configs}), or that their contrast, computed from their radius and temperature, is below the sensitivity of the instrument. The coldest detected companions have a temperature of 600~K.

The detections are globally in agreement with the T-type detection limit in \texttt{H-high} from Fig.~\ref{fig:SNR-ref-cases}, but some detected planets in the population are found below this curve. We note that this limit was averaged over ten companions at each separation and contrast. As seen in the lower right plot of Fig.~\ref{fig:CCFs-faint-H-high}, statistical variability puts some companions of a same sample above or below the 5-$\sigma$ threshold, and this might be the case for these companions in the population simulation. Another possible explanation is that the T-type detection limit was derived for companions at 800~K, and the planets below the curve in the population simulation have temperatures of 600 and 700~K, which might offer better contrast sensitivities.

Known directly imaged companions are all found outside the range of semi-major axes covered by HARMONI, showing the unprecedented observational window offered by the instrument. In terms of angular separation, only PDS~70~b falls inside the HARMONI field of view. A simple visual interpolation with the neighboring points tells us that it should be detected at a S/N of several hundreds. We compare in the left plot our results to the 10\% and 50\% detection probability curves of the VLT/SPHERE SHINE survey of 150 stars from \cite{Vigan2020}. Simulated companions in the $>10$\% and $>50$\% probability regions of SPHERE are respectively detected at S/N~>~48 and S/N~>~746 with HARMONI. HARMONI will thus allow to redetect and characterize known companions at very high S/N, as well as pushing much closer in and deeper the detection limits when looking for new companions around nearby stars.

We note that Eq. \eqref{eq:lum} is valid only if the irradiation by the host star is negligible compared to the intrinsic thermal emission of the companion \citep[see][Eqs. (1)-(8)]{Baraffe2003}. Irradiation contributes both to the thermal flux of the companion and the amount of light reflected by its surface, depending on the albedo, and is likely to affect the coldest and closest-in companions in our simulation. We note that on the detection side, as molecule mapping is based on cross-correlation with a model template, it could potentially remain efficient in presence of an additional minor reflected component. Nonetheless, accurate simulations for the coldest and closest-in companions would require to account for the stellar incident flux on the calculation of the atmospheric models, and to inject a reflected component taking into account a wavelength-dependent albedo, which is outside the scope of this study.

\section{Conclusions \& perspectives}
\label{sec:conclusions}

We have presented the first exoplanet detection limits of the high-contrast module of the future ELT/HARMONI instrument. We used an end-to-end model of the instrument simulating the observing conditions and adaptive optics correction, the high-contrast images, and the photometry of the observed objects. We then analyzed the simulated observations using the molecule mapping framework, based on the modeling and subtraction of the stellar and telluric components, and the cross-correlation of the residual datacube with a template spectrum. We used matched filtering to optimize the signal-to-noise of the cross-correlation signal.

Using this procedure, we have injected and attempted to detect T- and L-type companions around a F0 star with contrasts between 9 and 18~mag and separations between 50 and 300~mas, for the six different configurations of the high-contrast module of HARMONI. In the \texttt{HK} configuration at $R = 3500$, we detect companions with sensitivities $>5\sigma$ at contrasts up to 16~mag both for T- and L-type companions, with a flat sensitivity profile regarding to the angular separation. The \texttt{H} and \texttt{K} configurations at $R = 7000$, and \texttt{H-high} and \texttt{K1-high} at $R = 17\,000$, allow to detect T-type companions usually up to 15~mag, with a peak of sensitivity at 16~mag in the zone optimized by the apodizer, as close as 75~mas in $H$ band. The sensitivity to L types is decreased by $\sim$0.5~mag compared to T types in these configurations. In the \texttt{K2-high} configuration, L types are detected up to 15~mag and T types up to 13.5~mag.

We have applied the state-of-the-art angular differential imaging algorithm ANDROMEDA to the high-contrast images of HARMONI, and shown that molecule mapping is sensitive to contrasts up to $\sim$2.5~mag deeper than this technique. This confirms the high potential of medium-resolution spectro-imaging combined with dedicated post-processing techniques making full use of the spectral diversity.

We have simulated as well the influence of several input parameters such as the host star's spectral type and magnitude, and the seeing conditions. We find that the detection limits are not dependent on the spectral type of the host star. Increasing the host star brightness by 3~mag improves the detection limit only by 1~mag, whereas decreasing it by 3~mag lowers the detection limit by 3~mag as well. Regarding the seeing conditions, we find that sensitivities in the \texttt{HK} configuration are respectively lowered by $\sim$0.2 and $\sim$0.8~mag in the second and third quartiles of seeing conditions compared to the best one. This relatively low difference could allow to use HARMONI near optimal performances for exoplanet imaging during 60 to 70\% of telescope time at the ELT.

Finally, we have simulated the detection of planets from population models in order to better identify the reachable space of physical parameters. For a star located at 30~pc, we can detect all companions in the field of view of the instrument with semi-major axes $>2$~au and masses $>3~M_{\mathrm{Jup}}$, and partly detect them down to 1~au and $2~M_{\mathrm{Jup}}$. The $>10$\% and $>50$\% probability regions from the SHINE VLT/SPHERE survey of \cite{Vigan2020}, which include known directly imaged companions, are both located inside a very high-S/N zone with HARMONI. This shows that HARMONI will be able to reach populations of exoplanets currently inaccessible to current high-contrast imagers, such as VLT/SPHERE or Gemini/GPI, and redetect and characterize known companions at very high S/N.

HARMONI is expected to be available for the first light of ESO's ELT, currently planned in 2025. While not being an instrument specially designed for exoplanet imaging, its high-contrast module will allow to reach unprecedented separations and contrasts compared to current dedicated instruments. On the longer term, a dedicated high-contrast imager for the ELT, equipped with extreme adaptive optics and high spectral resolution, would allow to fully exploit the possibilities of the telescope and reach even fainter and closer-in exoplanets. One such instrument will be PCS \citep{Kasper2021}, whose ultimate goal is the characterization of potentially habitable rocky planets around nearby stars in the 2030s.

\begin{acknowledgements}
    We thank Dimitri Mawet for a useful discussion on matched filtering. This project has received funding from the European Research Council (ERC) under the European Union's Horizon 2020 research and innovation programme, grant agreement No. 757561 (HiRISE). This research has made use of computing facilities operated by CeSAM data center at LAM, Marseille, France.
\end{acknowledgements}

\bibliographystyle{aa}
\bibliography{paper}

\end{document}